\documentstyle[preprint,aps]{revtex}

\tighten

\begin{document}

\draft

\title{Conductance of a Single-mode Electron Waveguide\\
with Statistically Identical Rough Boundaries}

\author{N.~M.~Makarov, Yu.~V.~Tarasov}

\address{Institute for Radiophysics \& Electronics,
National Academy of Sciences of Ukraine,\\
12 Academician Proskura Str., Kharkov 310085, Ukraine}

\date{\today}

\maketitle

%------------------------------------------------------------------
\begin{abstract}
Transport characteristics of pure narrow $2D$ conductors, in which the
electron scattering is caused by rough side boundaries, have been
studied. The conductance of such strips is highly sensitive to the
intercorrelation properties of inhomogeneities of the opposite edges. The
case with completely correlated statistically identical boundaries (CCB)
is a peculiar one. Herein the electron scattering is uniquely due to
fluctuations of the asperity slope and is not related to the strip width
fluctuations. Owing to this, the electron relaxation lengths, specifically
the localization length, depend quite differently on the asperity
parameters as compared to the conductors with arbitrarily intercorrelated
edges. The method for calculating the dynamical characteristics of the CCB
electron waveguides is proposed clear of the restrictions on the asperity
height.
\end{abstract}
%-------------------------------------------------------------------


%------------------------------------------------------------------
\section{Introduction}
%------------------------------------------------------------------

Application of narrow conducting junctions with extremely small cross
dimensions in contemporary microelectronics has generated a great variety
of works on transport properties of such conductors. These properties were
proved to be substantially controlled by scattering of electrons at
random inhomogeneities of the conductor boundaries (see, e.g.,
Refs.~\cite{TrAsh,MakYur,KL,TakFer,Kun,KoKr,MeyStep,MakMor,LunKr} and
references therein). In particular, in Ref.~\cite{MakYur} pure single-mode
$2D$ conductors were shown to exhibit all peculiarities characteristic for
one-dimensional disordered systems. Their conductance is specified by the
coherent electron-surface scattering which causes the localization
effects. This certainly constrains lengthwise dimensions of narrow
microjunctions in view of the exponential increase of their resistivity
upon growing the length.

When producing $2D$ conductors of quite small width it is highly possible,
owing to the technology, for the opposite boundaries of the strips to have
exactly the same or sufficiently close statistical properties. Among all
the models of such statistically identical rough boundaries two
substantially different are distinguished. One of them includes the
strips with no correlation between the asperities of the opposite edges.
Within the other model, correlation between the asperities of the opposite
boundaries is just the same as the correlation at any strip edge.
Boundaries of the latter type will be referred to as completely correlated
(CCB). In Ref.~\cite{MakYur} the electron scattering caused by
irregularities of only one boundary of the conducting strip was analyzed,
the other being perfectly smooth. The obtained results are clearly
applicable for the former (not intercorrelated) kind of boundaries. At the
same time, the CCB conductors have not received due attention so far.

In this contribution, the CCB case is examined and shown to be the special
one. The model considered is physically equivalent to that when the
conductor width keeps constant (or nearly constant) along the whole
length, despite inhomogeneities of the strip edges. The local mode
structure of the electron waveguide remains therein undisturbed. As a
result, the electron scattering is due not to the asperity heights, whose
values are not restricted in the problem, but to the asperity slopes only.

It is well known that the by-height scattering is controlled by the
parameter $(k_F\sigma)^2$ ($k_F$ the Fermi wavenumber of electrons), and
the electron relaxation rate is proportional to the square of the r.m.s.
asperity height $\sigma$ (see, e.g., Refs.~\cite{MakYur,MeyStep}). We
argue below that in the single-mode CCB strips the main controlling factor
is the ratio $(\sigma/R_c)^4$ ($R_c$ the correlation radius of the boundary
asperities). Therefore, the electron scattering rate is proportional to
higher, namely the fourth, power of $\sigma$. At first glance it should
give rise to an increase of the localization length as compared to that
from Ref.~\cite{MakYur}. However, it is not the case as a rule. In a
single-mode CCB strip even with mildly sloping boundary asperities the
electron localization length at certain, easily reachable, conditions
appears to be much less than the by-height scattering length.

%--------------------------------------------------------------------
\section{Formulation of the problem}
%--------------------------------------------------------------------

Let a two-dimensional conducting strip of the length $L$ and
the average width $d$ occupy the region of $(x,z)$ plane specified by the
inequalities

\begin{equation}
-L/2 \leq x \leq L/2,  \qquad \qquad
\xi_1(x) \leq z \leq d+\xi_2(x).
\end{equation}
The functions $\xi_{1,2}(x)$ describe asperities of the edges of the
strip. We assume them continuously differentiable random processes with
zero mean values. The correlation properties thereof will be thoroughly
discussed below.

In accordance with the standard linear response theory \cite{Kubo}
conductance (as well as conductivity) is expressed through product of
differences between the advanced and retarded one-electron Green functions
(see, e.g., Refs.~\cite{FishLee,Edwards}). In what follows the
electron scattering will be supposed weak (see Eq.~(\ref{two-scale})). It
is well-proved \cite{Abrik,Efetov} that under these conditions one can
neglect the products of the like Green functions (both retarded and both
advanced) in the general expression for the conductance. Taking into
account the relation between the advanced and retarded Green functions,
the conductance $G(L)$ of the strip, divided by the conductance quantum
$e^2/\pi\hbar$, at zero temperature is represented as

\begin{equation}
\frac{G(L)}{e^2/\pi\hbar} =
- \frac{4}{L^2} \int_{-L/2}^{L/2}dx
\int_{\xi_1(x)}^{d+\xi_2(x)}dz \;
\int_{-L/2}^{L/2}dx^\prime
\int_{\xi_1(x^{\prime})}^{d+\xi_2(x^\prime)}dz^\prime
\frac{\partial{\cal G}(x,x^\prime;z,z^\prime)}{\partial x}
\frac{\partial{\cal G}^* (x,x^\prime;z,z^\prime)}{\partial x^\prime} \, .
\label{N(L)}
\end{equation}
Here ${\cal G}(x,x^\prime;z,z^\prime)$ is the retarded one-electron Green
function obeying the equation

\begin{equation}
\left[ {\partial^2 \over \partial x^2}+{\partial^2 \over \partial z^2}+
(k_F+i0)^2 \right]{\cal G}(x,x^\prime;z,z^\prime) =
\delta(x-x^\prime)\delta(z-z^\prime)
\label{G-eq1}
\end{equation}
with $k_F$ the Fermi wavenumber. The asterisk in Eq.~(\ref{N(L)}) denotes
complex conjugation. We consider the function ${\cal G}$ meeting zero
Dirichlet boundary conditions at the strip edges $z=\xi_1(x)$ and
$z=d+\xi_2(x)$ whereas at the strip ends $x=\pm L/2$ the radiative
conditions are satisfied.

In solving problems related to the boundary scattering in waveguides the
coordinate transformation is often applied to smooth out both boundaries
toward ideally flat (see, e.g., Ref.~\cite{MeyStep}). For our purpose it
is more conveniently to smooth only one side of the strip. Let it be, for
definiteness, the lower one which we smooth out to the line $z_{new}=0$.
This is done by a transformation of the transverse coordinate,
$z_{new}=z_{old}-\xi_1(x)$, accompanied by the corresponding change of the
longitudinal velocity operator. As a result, the perturbation $\xi_1(x)$
is transferred to both the conductance expression (\ref{N(L)}),

\begin{eqnarray}
\frac{G(L)}{e^2/\pi\hbar} &=& - \frac{4}{L^2}
\int_{-L/2}^{L/2}dx \int_0^{d(x)}dz \;
\int_{-L/2}^{L/2}dx^\prime \int_0^{d(x')} dz^\prime \times
\nonumber \\ \cr
&\times&
\left[\frac{\partial}{\partial x} -
\xi_1'(x)\frac{\partial}{\partial z} \right]
{\cal G}(x,x^\prime;z,z^\prime)
\left[\frac{\partial}{\partial x^\prime} -
\xi_1'(x^\prime)\frac{\partial}{\partial z^\prime} \right]
{\cal G}^*(x,x^\prime;z,z^\prime) \, ,
\label{N-main}
\end{eqnarray}
and to the Green function equation (\ref{G-eq1}) which takes the form

\begin{eqnarray}
\left[ {\partial^2 \over \partial x^2}\right. &+& \left.
\alpha^2 {\partial^2 \over \partial z^2}+
(k_F+i0)^2 \right] {\cal G}(x,x^\prime;z,z^\prime) - \nonumber \\ \cr
&-& \left[ \hat{\cal U}(x) {\partial \over \partial z} -
\hat{\cal V}(x) {\partial^2 \over \partial z^2} \right]
{\cal G}(x,x^\prime;z,z^\prime) = \delta(x-x^\prime)\delta(z-z^\prime) \ .
\label{G-eq2}
\end{eqnarray}
From here on we use the notations listed below. In Eq.~(\ref{N-main})
$d(x)$ stands for the local width of the strip,

\begin{equation}
d(x) = d + \Delta \xi(x), \qquad \qquad
\Delta \xi(x) = \xi_2(x) - \xi_1(x) \ ,
\label{d(x)}
\end{equation}
with $\Delta \xi(x)$ being the width fluctuation. Next, in
Eq.~(\ref{G-eq2}) the factor $\alpha^2$ and the effective zero-mean-valued
`potentials' $\hat{\cal V}(x)$ and $\hat{\cal U}(x)$ of the
electron-surface interaction  have been introduced,

\begin{equation}
\alpha^2 = 1 + \langle {\xi'_1}^2(x) \rangle,
\qquad
\hat{\cal V}(x) = {\xi'_1}^2(x) - \langle {\xi'_1}^2(x) \rangle,
\qquad
\hat{\cal U}(x) =
\xi_1'(x)\frac{\partial}{\partial x} +
\frac{\partial}{\partial x}\xi_1'(x).
\label{VUdef}
\end{equation}
The angular brackets $\langle \ldots \rangle$ denote averaging over
realizations of the random functions $\xi_{1,2}(x)$, primes in functions
stand for derivatives over their arguments.

To analyze the electron transport in a narrow $2D$ waveguide, where
quantization of the electron transverse motion is rather considerable, we
apply the discrete, i.e. `mode', representation in the coordinate $z$.
The Green function now turns to zero at $z=0$ and $z=d(x)$. Then, allowing
for this, we present ${\cal G}(x,x';z,z')$ as a series,

\begin{equation}
{\cal G}(x,x';z,z') = {2 \over \sqrt{d(x)d(x')}}
\sum\limits_{n,n'=1}^\infty G_{nn'}(x,x')
\sin\left( {\pi n z \over d(x)} \right)
\sin\left( {\pi n' z' \over d(x')} \right) \, .
\label{G-ser}
\end{equation}
By substituting Eq.~(\ref{G-ser}) into Eq.~(\ref{G-eq2}) we arrive at the
following set of equations for the Fourier coefficients $G_{nn'}(x,x')$,

\begin{eqnarray}
&&\left\{ {\partial^2 \over \partial x^2} +
k_n^2(x) + i0 - \left[ {\pi n \over d(x)} \right]^2
\hat{\cal V}(x) \right\}
G_{nn'}(x,x') - {4 \over d(x)} \sum\limits_{m=1}^\infty
A_{nm} \hat{\cal U}(x)G_{mn'}(x,x') + \nonumber \\
&&+ {2 \over d(x)} \sum\limits_{m=1}^{\infty}
\hat\Phi_{nm}(x)G_{mn'}(x,x') =
\delta_{nn'}\delta(x-x') \, .
\label{Gnn'1}
\end{eqnarray}
Here the locally quantized value $k_n(x)$ of the electron longitudinal
wavenumber and the coefficient matrix $A_{nm}$ are given by

\begin{equation}
k_n(x) = \left(k_F^2 - \left[{\pi n \alpha \over
d(x)}\right]^2\right)^{1/2},
\qquad
A_{nm} = {nm \over n^2-m^2} \sin^2\left[{\pi \over 2}(n-m)\right].
\label{k_n}
\end{equation}
We omit the expression for the matrix potential $\hat\Phi_{nm}(x)$ in
view of its awkwardness. It is only important for us to point out its
being the functional of $\xi_1'(x)$ and $\Delta \xi'(x)$ and turning to
zero as $\Delta \xi'(x)=0$.

The equation (\ref{Gnn'1}) covers scattering of electrons by rough
boundaries of two-dimensional electron waveguide at arbitrary correlation
conditions for the asperity heights $\xi_{1,2}(x)$. In this work, our
intention is to discuss the case of not arbitrary but statistically
identical strip sides. Moreover, we deal with the conductors where
asperities of the opposite sides correlate with each other just as they
do within every edge of the strip. For stating this CCB model of $2D$
junction we use the correlation equalities

\begin{equation}
\langle \xi_{i}(x) \rangle = 0; \qquad \qquad
\langle \xi_{i}(x)\xi_{k}(x') \rangle = \sigma^2 {\cal W}(x-x'), \qquad
i,k=1,2 .
\label{xi-corr}
\end{equation}
Here ${\cal W}(x)$ is the correlation coefficient specified by the unity
amplitude and the correlation radius $R_c$. As a consequence of
Eq.~(\ref{xi-corr}), the following correlation functions equal zero,

\begin{equation}
\langle \xi_{1,2}(x)\Delta\xi(x') \rangle =
\langle \Delta\xi(x)\Delta\xi(x') \rangle = 0 \ .
\label{corr=0}
\end{equation}
For the weak electron-surface scattering (or gaussian statistics of
the asperities), Eq.~(\ref{corr=0}) leads to the same result for any
averaged quantity as at $\Delta\xi(x) = \Delta\xi'(x) = 0$. So hereinafter
the local width of the strip, $d(x)$, can be replaced by its average value
$d$, and the last term containing the potential $\hat\Phi_{nm}(x)$ in
l.h.s. of Eq.~(\ref{Gnn'1}) can be properly dropped. Below we omit the
subscript `1' on the function $\xi_1(x)$ for simplicity.

Deviation of the factor $\alpha^2$ from unity in $k_n$, Eq.~(\ref{k_n}),
could be significant at `sharp' asperities as it causes effective decrease
of number of the modes propagating in the waveguide. Taking this into
account is of no crucial problem. Nevertheless, we introduce one more
simplification not to complicate calculations. We will consider only the
mildly sloping boundary inhomogeneities for which

\begin{equation}
|\xi'_{1,2}(x)|^2 \ll 1 \ .
\label{smooth}
\end{equation}
This allows to put henceforward  $\alpha^2 = 1$ and neglect perturbation
of the velocity operators in the expression (\ref{N-main}) for the
conductance.

Note that in Eq.~(\ref{Gnn'1}) the term containing the potential
$\hat{\cal V}(x)$ describes the intrachannel (intramode) electron
scattering with conservation of the quantum number $n$. At the same
time, the perturbation operator $\hat{\cal U}(x)$ results, to the
basic approximation, just in the intermode scattering since the
corresponding sum over $m$ in Eq.~(\ref{Gnn'1}) is free of the term with
$m=n$ ($A_{nn}=0$, in accordance with a definition from Eq.~(\ref{k_n})).
The inverse lengths of the electron scattering from the potentials
$\hat{\cal V}(x)$ and $\hat{\cal U}(x)$ are proportional, in the main
approximation, to $\langle \xi'^4(x) \rangle$ and
$\langle \xi'^2(x)\rangle$, respectively. If the boundary asperities are
mildly sloping (\ref{smooth}), these lengths could substantially differ.
However, in the case of narrow conductors with a single propagating
electron mode (the ultra-quantum limit), when

\begin{equation}
1 < k_Fd/\pi <2,
\label{1-chan}
\end{equation}
the term linear in the operator $\hat{\cal U}(x)$ multiplied by
$G_{11}(x,x')$ is not present in Eq.~(\ref{Gnn'1}). That is why the
spatial decrease of the average single-mode Green function
$<G_{11}(x,x')>$ is not determined by the interchannel but the
intrachannel electron scattering with the attenuation length proportional
to $\sigma^{-4}$. This is just the case we analyze below.

For the benefit of our study an important point is to presume the
electron-surface scattering weak. That is the electron relaxation length
$L_1$ in the open channel with $n=1$ has to be large as compared to
`microscopic' lengths of our problem, specifically the electron wavelength
$k_1^{-1}$ and the correlation radius $R_c$. What is more, the conductor
length $L$ will be supposed obeying the similar requirements, which are
necessary for averaging procedure to be reasonable. All these conditions
can be formulated through the inequality

\begin{equation}
\max\{k_1^{-1},R_c\} \ll \min\{L_1,L\} \, .
\label{two-scale}
\end{equation}
Note that we do not assume any predetermined interrelation between $L$ and
$L_1$ as well as between $k_1^{-1}$ and $R_c$.

To get the starting expression for the single-mode conductance $G_1(L)$
one should substitute Eq.~(\ref{G-ser}) into Eq.~(\ref{N-main}). In line
with the weak-scattering conditions (\ref{two-scale}), all the Green
functions with $n,n'\neq 1$ contribute $G_1(L)$ slightly. Then for the
dimensionless single-mode conductance $T_1(L)$ we have

\begin{equation}
T_1(L) = \frac{G_1(L)}{e^2/\pi\hbar} = -{4 \over L^2}
\int\!\!\!\int_{-L/2}^{L/2} dxdx'
{\partial G_{11}(x,x') \over \partial x}
{\partial G_{11}^*(x,x') \over \partial x'}.
\label{N-1-def}
\end{equation}

As it was pointed out, the equation (\ref{Gnn'1}) with $n=n'=1$ does
not contain the first degree of the potential $\hat{\cal U}(x)$ at the
function $G_{11}(x,x')$. For this reason in a single-channel strip the
electron-surface scattering caused by the potential $\hat{\cal U}(x)$
manifests itself in higher orders of its magnitude. To obtain the
correct equation for $G_{11}(x,x')$ one has to follow the procedure
outlined in Appendix~\ref{G11-K}. In the event of mildly sloping
asperities (\ref{smooth}) and weak-scattering approximation
(\ref{two-scale}) we get

\begin{eqnarray}
&&\left( {\partial^2 \over \partial x^2} + k_1^2 + i0 \right)
G_{11}(x,x') - \left(\pi \over d \right)^2
\hat{\cal V}(x)G_{11}(x,x') -
\nonumber \\ \cr
&& - \left(4 \over d \right)^2
\int_{-L/2}^{L/2} dx_1 \, {\hat K}(x,x_1)G_{11}(x_1,x')
= \delta(x-x')\, .
\label{G11-eq}
\end{eqnarray}
Here the novel perturbation operator has occurred with the kernel

\begin{equation}
{\hat K}(x,x') = - \sum_{m=2}^{\infty}
A_{1m}^2 \left[ \hat{\cal U}(x) G_m^{(0)}(|x-x'|) \hat{\cal U}(x') -
\langle \hat{\cal U}(x) G_m^{(0)}(|x-x'|) \hat{\cal U}(x') \rangle
\right].
\label{K-ker}
\end{equation}
The unperturbed Green functions $G_m^{(0)}(|x-x'|)$ of the modes
$m \geq 2$ attenuate exponentially along the strip over the electron
wavelengths,

\begin{equation}
G_m^{(0)}(|x-x'|) = - {1 \over 2|k_m| }\exp\Bigl( -|k_m||x-x'|\Bigr),
\qquad
|k_m| = \left[ (\pi m/d)^2-k_F^2\right]^{1/2}  \ .
\label{G_m0}
\end{equation}

Thus, the problem is reduced to calculating the statistical moments
$\langle T_1^n(L)\rangle$ of the conductance (\ref{N-1-def}) with the
single-mode Green functions found from Eq.~(\ref{G11-eq}).

%----------------------------------------------------------
\section{Two-scale model}
\label{Model}
%-----------------------------------------------------------

The equation (\ref{G11-eq}) for the Green function $G_{11}(x,x')$ is
strictly one-dimensional and, consequently, makes it possible to analyze
in detail the effects of coherent multiple scattering of electrons.
Inhomogeneities of the strip edges enter now the scattering potentials of
the equation rather than the boundary conditions for the Green functions.
In accordance with the weak-scattering assumption (\ref{two-scale}), there
exist two groups of substantially different spatial scales in our problem.
On the one hand, it is a group of `macroscopic' lengths, $L_1$ and $L$,
and on the other a pair of the `microscopic' lengths, $k_1^{-1}$ and
$R_c$. This suggests that it is reasonable to apply for calculating the
Green function $G_{11}$ the two-scale model of oscillations.

Take the well-known representation for the one-dimensional Green function
$G_{11}(x,x')$,

\begin{equation}
G_{11}(x,x') = \widetilde{W}^{-1} \big[ \psi_+(x)\psi_-(x')\Theta(x-x') +
\psi_+(x')\psi_-(x)\Theta(x'-x) \big] \ .
\label{G-psi}
\end{equation}
In Eq.~(\ref{G-psi}), the functions $\psi_{\pm}(x)$ are the linearly
independent solutions of the uniform equation (\ref{G11-eq}) with the
radiation conditions satisfied at the strip ends $x=\pm L/2$,
respectively. The Wronskian of those functions is $\widetilde{W}$, and
$\Theta(x)$ is the Heaviside unit-step function. The functions
$\psi_{\pm}(x)$ will be sought as superposition of modulated waves
propagating in opposite directions along the $x$-axis,

\begin{equation}
\psi_{\pm}(x) = \pi_{\pm}(x)\exp(\pm ik_1 x) -
i\gamma_{\pm}(x)\exp(\mp ik_1 x) \ .
\label{psi-pm}
\end{equation}
The radiation conditions for the functions $\psi_{\pm}(x)$ are stated as
the `initial' conditions for the amplitudes $\pi_{\pm}(x)$ and
$\gamma_{\pm}(x)$, i.e.

\begin{equation}
\pi_{\pm}(\pm L/2) = 1 \ , \qquad \qquad \gamma_{\pm}(\pm L/2) = 0 \ .
\label{Bcond}
\end{equation}
Emphasize that the amplitudes $\pi_{\pm}(x)$ and $\gamma_{\pm}(x)$ in
Eqs.~(\ref{psi-pm}), (\ref{Bcond}) are varied at the characteristic
length $L_1$ (or $L$). Therefore in the framework of two-scale
approximation (\ref{two-scale}) they are smooth functions of $x$ as
compared to the rapidly oscillating exponents $\exp(\pm ik_1 x)$ and the
correlation coefficient ${\cal W}(x)$.

According to Eqs.~(\ref{G-psi}), (\ref{psi-pm}), the problem of
calculating the Green function $G_{11}$ is reduced to finding the smooth
amplitudes $\pi_{\pm}(x)$ and $\gamma_{\pm}(x)$. Within the
assumption (\ref{two-scale}), the appropriate equations for them are
deduced by the standard method of averaging over the rapid phases (see,
e.g., Ref.~\cite{Bogol}). For doing that one should substitute
$\psi_{\pm}(x)$ of the form (\ref{psi-pm}) into the uniform equation
(\ref{G11-eq}) and multiply it by $\exp(\mp i k_1 x)$. Then the equation
obtained should be averaged over the spatial interval of a length
intermediate between the above introduced macroscopic and microscopic
scales. The same should be done using the multiplier $\exp(\pm i k_1 x)$.
As a result, we get the set of dynamic equations,

\begin{eqnarray}
&&\pi'_{\pm}(x)\pm i\eta(x)\pi_{\pm}(x)
\pm\zeta_{\pm}^*(x)\gamma_{\pm}(x)=0 \ ,
\nonumber \\
&& \label{PiGam} \\
&&\gamma'_{\pm}(x)\mp i\eta(x)\gamma_{\pm}(x)
\pm\zeta_{\pm}(x)\pi_{\pm}(x)=0 \ .
\nonumber
\end{eqnarray}
The variable coefficients $\eta(x)$ and $\zeta_{\pm}(x)$ are the
space-averaged random fields associated with the electron-surface
interaction potentials from Eq.~(\ref{G11-eq}). The function $\eta(x)$ is
a real field whereas $\zeta_{\pm}(x)$ are the complex conjugated ones.
Since our concern is with the quantities averaged over realizations of the
random function $\xi(x)$, only the correlation properties of the fields are
of decisive importance. In the Appendix~\ref{Fields} the exact expressions
for $\eta(x)$ and $\zeta_{\pm}(x)$ are written and it is shown that within
the two-scale model (\ref{two-scale}) all these functions can be properly
regarded as $\delta$-correlated gaussian random processes with the
correlation relations as follows,

\begin{eqnarray}
&\langle \eta(x) \rangle = \langle \zeta_{\pm}(x) \rangle =
\langle \eta(x)\zeta_{\pm}(x') \rangle =
\langle \zeta_{\pm}(x)\zeta_{\pm}(x') \rangle = 0 \ , & \nonumber \\
&& \label{etazeta-corr}\\
&\langle \eta(x)\eta(x') \rangle = {L_f}^{-1}\delta(x-x') \ ,
\quad
\langle \zeta_{\pm}(x)\zeta_{\pm}^*(x')  \rangle =
{L_b}^{-1}\delta(x-x')& \ . \nonumber
\end{eqnarray}
Here in Eq.~(\ref{etazeta-corr}) two lengths are present, $L_f$ and $L_b$,
specified by the expressions

\begin{eqnarray}
{L_f}^{-1} &=& {1 \over 2k_1^2}\left({\pi\sigma \over d}\right)^4
\int_{-\infty}^\infty {dq_x \over 2\pi}\, q_x^4 W^2(q_x)
\times \nonumber \\ \cr
&\times& \left\{ 1+{8\over\pi^2}\sum\limits_{m'=2}^\infty
A_{1m'}^2\left[ (2k_1+q_x)^2 g_{m'}^{(0)}(k_1+q_x) +
(2k_1-q_x)^2 g_{m'}^{(0)}(k_1-q_x) \right]\right\}^2 \ ,
\label{L_f}
\end{eqnarray}
\begin{eqnarray}
{L_b}^{-1} &=& {1 \over 2k_1^2}\left({\pi\sigma \over d}\right)^4
\int_{-\infty}^\infty {dq_x \over 2\pi}\,
(q_x^2-k_1^2)^2 W(q_x-k_1)W(q_x+k_1) \times \nonumber \\ \cr
&\times& \left[ 1+\left({4\over\pi}\right)^2
\sum\limits_{m=2}^\infty A_{1m}^2(q_x^2-k_1^2) g_m^{(0)}(q_x)\right]^2 \ .
\label{L_b}
\end{eqnarray}
The function $W(q_x)$ is the Fourier transform of the correlation
coefficient ${\cal W}(x)$ from Eq.~(\ref{xi-corr}), and $g_m^{(0)}(q_x)$
is the analogous transform of the unperturbed Green function~(\ref{G_m0}),

\begin{equation}
g_m^{(0)}(q_x) = -{1 \over q_x^2+|k_m|^2} \ .
\label{g-m0}
\end{equation}

Taking advantage of Eqs.~(\ref{G-psi}), (\ref{psi-pm}), and
(\ref{etazeta-corr}) we can show that superposition of the inverse
lengths (\ref{L_f}) and (\ref{L_b}) is the inverse outgoing length of
attenuation of the average Green function $\langle G_{11}(x,x') \rangle$.
It is reasonable then to associate this superposition with the length
$L_1$ from Eq.~(\ref{two-scale}), i.e.

\begin{equation}
L_1^{-1} = L_f^{-1} + L_b^{-1} \, .
\label{L1LfLb}
\end{equation}
From the derivation presented in Appendix \ref{Fields}, as well as from
the appearance itself of the expressions (\ref{L_f}) and (\ref{L_b}), it
is easy to establish that the length $L_f$ is related to the forward
electron scattering (i.e. without changing the sign of the velocity
$x$-component) while $L_b$ to the backward scattering. In our
consideration the length $L_f$ specifies the correlator
$\langle \eta(x)\eta(x') \rangle$ whereas $L_b$ controls the correlator
$\langle \zeta_{\pm}(x)\zeta_{\pm}^*(x')  \rangle$. Hence the
conclusion is clear that the fields $\eta(x)$ and $\zeta_{\pm}(x)$ from
Eq.~(\ref{PiGam}) are responsible for the forward and backward electron
scattering, respectively.

%------------------------------------------------------------
\section{Conductance and resistivity moments}
%-------------------------------------------------------------

The next step is to express the dimensionless conductance (\ref{N-1-def})
through the smooth amplitudes $\pi_{\pm}$ and $\gamma_{\pm}$ and to
average it subsequently over the random fields $\eta(x)$ and
$\zeta_{\pm}(x)$. To do this substitute Eqs.~(\ref{G-psi}),
(\ref{psi-pm}) into Eq.~(\ref{N-1-def}). After a succession of simple
transformations with the use of the inequalities (\ref{two-scale}) we get
the formula for the conductance of a single-mode strip,

\begin{equation}
T_1(L) =
|\pi_{\pm}^{-1}(\mp L/2)|^2 \ .
\label{N<->t_L}
\end{equation}
From this equality it naturally follows that the quantity
$\pi_{\pm}^{-1}(\mp L/2)$ can be regarded as the amplitude transmission
coefficient of the waveguide of the length $L$.

Introduce the amplitude reflection coefficient
$\Gamma_{\pm}(x)=\gamma_{\pm}(x)/\pi_{\pm}(x)$, in accordance with
Eq.~(\ref{psi-pm}). From Eq.~(\ref{PiGam}) it can be established that the
quantities $\pi_{\pm}^{-1}(x)$ and $\Gamma_{\pm}(x)$, in line with their
physical meaning, obey the flow conservation law,

\begin{equation}
|\Gamma_{\pm}(x)|^2 + |\pi_{\pm}^{-1}(x)|^2 =1.
\label{Potok}
\end{equation}
As a consequence of Eqs.~(\ref{PiGam}), (\ref{Bcond}), the coefficient
$\Gamma_{\pm}(x)$ satisfies the Riccati-type equation with the homogeneous
initial condition,

\begin{eqnarray}
\pm {d\Gamma_{\pm}(x) \over dx} &=& 2i\eta(x)\Gamma_{\pm}(x) +
\zeta_{\pm}^*(x)\Gamma_{\pm}^2(x) - \zeta_{\pm}(x) \ ,
\label{Gam-eq} \\   \cr
&&\Gamma_{\pm}(\pm L/2) = 0 \ .
\nonumber
\end{eqnarray}
Being closed, this equation is more convenient to analyze than the set
(\ref{PiGam}). Therefore, expressing the single-mode conductance
(\ref{N<->t_L}) through $|\Gamma_{\pm}(\mp L/2)|^2$ by the use of the
conservation law (\ref{Potok}), we will perform all the following
calculations in terms of the reflection coefficient $\Gamma_{\pm}(x)$
rather than the transmission one $\pi_{\pm}^{-1}(x)$.

Attention should be given to the fact that the field $\eta(x)$ may be
eliminated from Eq.~(\ref{Gam-eq}) by concurrent phase transformations of
the reflection coefficient $\Gamma_{\pm}(x)$ and the fields
$\zeta_{\pm}(x)$. These transformations retain the correlation relations
(\ref{etazeta-corr}) for the new renormalized fields $\zeta_{\pm}(x)$
unaffected. That is one can put the random function $\eta(x)$ in
Eq.~(\ref{Gam-eq}) equal to zero. Consequently, the outcome for arbitrary
moment of the conductance is specified by just backscattering of
electrons, i.e. by the attenuation length $L_b$ from Eq.~(\ref{L_b}).

Now let us define the $n$-th moment of the reflection coefficient squared
modulus,

\begin{equation}
R_n^{\pm}(x)=\langle |\Gamma_{\pm}(x)|^{2n}\rangle \ .
\label{R_n-def}
\end{equation}
From Eq.~(\ref{Gam-eq}), one can obtain, basing on the Furutsu-Novikov
formula and the correlation relations (\ref{etazeta-corr}), the
differential-difference equation for that moment (see, e.g.,
Ref.~\cite{Papanic}),

\begin{equation}
\pm{dR_n^{\pm}(x) \over dx} = -{n^2\over  L_b}
\left[ R_{n+1}^{\pm}(x)-2R_n^{\pm}(x)+R_{n-1}^{\pm}(x)\right],
\qquad n=0,1,2,\ldots \, ,
\label{R_n-eq}
\end{equation}
with the initial condition on the coordinate $x$

\begin{equation}
R_n^{\pm}(\pm L/2) = \delta_{n0}.
\label{R_n-cond}
\end{equation}
Besides the condition (\ref{R_n-cond}), we have $R_0^{\pm}(x)=1$ and
$R_n^\pm(x) \to 0$ as $n\to\infty$, in accordance with the
definition (\ref{R_n-def}).

Solution of Eq.~(\ref{R_n-eq}) that matches all the above conditions can
be expressed through the distribution function $P_L^\pm(u,x)$ and, upon
due parametrization, represented as

\begin{equation}
R_n^{\pm}(x) = \int_1^\infty du P_L^\pm(u,x)
\left({u-1 \over u+1}\right)^n.
\label{R_n-int}
\end{equation}
In line with this representation, statistical moments of the conductance
(\ref{N<->t_L}) can be written through the same distribution function,

\begin{equation}
\langle T_1^n(L)\rangle = \langle (1-|\Gamma_{\pm}(\mp L/2)|^2)^n\rangle
= \int_1^\infty du P_L^\pm(u,\mp L/2)
\left({2 \over u+1}\right)^n.
\label{T_n-int}
\end{equation}
So just the probability density $P_L^\pm(u,x)$ is of our need.

Substitute $R_n^{\pm}(x)$ in the form (\ref{R_n-int}) into equation
(\ref{R_n-eq}) and perform some elementary transformations. Then we get
for $P_L^\pm(u,x)$ the Fokker-Plank equation

\begin{equation}
\pm L_b{\partial P_L^\pm(u,x) \over \partial x} =
-{\partial \over \partial u}(u^2-1)
{\partial P_L^\pm(u,x) \over \partial u},
\label{Fokk-Plank}
\end{equation}
which is supplemented, according to Eq.~(\ref{R_n-cond}), by the
initial conditions on the coordinate $x$,

\begin{equation}
P_L^\pm(u,\pm L/2) = \delta(u-1-0) \ .
\label{P_L-ic}
\end{equation}
From the equality $R_0^{\pm}(x)=1$ normalization of the function
$P_L^\pm(u,x)$ to unity follows. In its turn, this implies the
distribution function to be integrable over the variable $u$, in
particular, at $u\to 1$ and $u\to\infty$.

The solution of Eq.~(\ref{Fokk-Plank}), which satisfies the above pointed
requirements, is well-established (see, e.g., Ref.~\cite{LifGredPas}). It
can be obtained by the use of the Mehler-Fock transformation
\cite{Mehler,Fock} and found to have the conventional form

\begin{eqnarray}
P_L^\pm(\cosh\alpha,x) &=& {1 \over \sqrt{8\pi}}
\left( {L \mp 2x \over 2L_b} \right)^{-3/2}
\exp \left(-{L \mp 2x \over 8L_b} \right) \times
\nonumber \\ \cr
&\times&
\int_{\alpha}^{\infty} {v\,dv \over (\cosh v - \cosh\alpha)^{1/2}}
\exp \left[ -{v^2 \over 4} \left({L \mp 2x \over 2L_b} \right)^{-1}
\right] \ ,
\label{P_L} \\ \cr
&& u = \cosh\alpha, \qquad  \alpha \ge 0.
\nonumber
\end{eqnarray}
With this solution we get from Eq.~(\ref{T_n-int}) a relatively simple,
as well suitable to analyze, expression for the $n$-th moment of the
dimensionless conductance $T_1(L)$,

\begin{eqnarray}
\langle T_1^n(L)\rangle &=&
{4\over\sqrt\pi} \left({L_b\over L}\right)^{3/2}
\exp{\left(-{L\over 4L_b}\right)} \times \nonumber \\
&\times& \int_0^\infty{zdz \over \cosh^{2n-1}z}
\exp\left(-z^2{L_b\over L}\right)
\int_0^z dy\, \cosh^{2(n-1)}y \, , \qquad n=0,\pm 1,\pm2,\ldots \ .
\label{N-mom}
\end{eqnarray}
The formula (\ref{N-mom}) completely determines the main averaged
transport characteristics of a single-mode conducting strip.

%----------------------------------------------------------------
\section{Results and discussion}
%----------------------------------------------------------------

Let us write down the expressions for the average dimensionless
conductance $\langle T_1(L) \rangle$ and resistance
$\langle T_1^{-1}(L) \rangle$. Put $n=1$ in Eq.~(\ref{N-mom}) and
take the integrals asymptotically in the parameter $L/L_b$. Then the
asymptotic expressions for the average conductance look like

\begin{eqnarray}
\langle T_1(L) \rangle \approx 1-L/L_b  \qquad
&\mbox{ if }& \qquad L/L_b\ll 1 \ , \nonumber \\
\label{N1-ass} \\
\langle T_1(L) \rangle \approx
2^{-1}\pi^{5/2}\left(L/L_b\right)^{-3/2}
\exp\left(-L/ 4L_b\right) \qquad
&\mbox{ if }& \qquad L/L_b\gg 1 \ .
\nonumber
\end{eqnarray}
At $n=-1$ all the integrals are calculated exactly in Eq.~(\ref{N-mom}),
and for the average dimensionless resistance we get the formula,

\begin{equation}
\langle T_1^{-1}(L) \rangle =
{1 \over 2}\left[1+\exp\left({2L\over L_b}\right)\right] \ .
\label{sopr}
\end{equation}
For the sake of completeness, we also give, without proof, the averaged
logarithm of the dimensionless conductance,

\begin{equation}
\langle \ln T_1(L) \rangle = - L/L_b \ .
\label{lnN_1}
\end{equation}
It can be found directly from the equations (\ref{PiGam}).

The results (\ref{N1-ass}) -- (\ref{lnN_1}) match absolutely the concepts
of the localization theory for one-dimensional disordered conductors and
therefore coincide in appearance with those obtained, in particular, in
Ref.~\cite{MakYur}. The asymptotics (\ref{N1-ass}) show exponential
decrease of the average conductance as the strip length $L$ exceeds the
localization length $L_{loc}=4L_b$. The expression (\ref{sopr}) describes
exponential growth of the average resistance with growing the strip length
$L$. Needless to say that both the conductance and the resistance are not
self-averaged quantities. The main difference of our results from the
previously obtained is in the relaxation length $L_b$, Eq.~(\ref{L_b}), to
be discussed below.

A few words about the validity range for the results (\ref{L_f}),
(\ref{L_b}), (\ref{N-mom}) -- (\ref{lnN_1}). First of all, the boundary
asperities of the electron waveguide were supposed to be mildly sloping.
The corresponding requirement (\ref{smooth}) sets limits on the relation
between the asperity height and length,

\begin{equation}
(\sigma/R_c)^2 \ll 1 \ .
\label{Smooth-2}
\end{equation}
Additional restrictions result from the weakness of the
electron-surface scattering, Eq.~(\ref{two-scale}). In accordance with
Eq.~(\ref{L1LfLb}), the length $L_f$ can be used therein as the
parameter $L_1$, since the inequality $L_f \alt L_b$ always holds true.
One of the conditions (\ref{two-scale}), namely, $L_1 \gg R_c$, is
reduced to smallness of the Fresnel parameter $k_F\sigma^2/R_c$. In terms
of the diffraction theory, it means the absence of the shadowing effect in
scattering of the electron waves by rough boundaries (see, e.g.,
Ref.~\cite{BassFuks}). This condition can be rewritten via the parameters
of our problem as follows,

\begin{equation}
\sigma^2 / R_c d \ll 1 \ .
\label{Frenel}
\end{equation}
The second inequality from Eq.~(\ref{two-scale}), $L_1 \gg k_1^{-1}$, is
reduced merely to the product of Eq.~(\ref{Smooth-2}) and
Eq.~(\ref{Frenel}), so it holds automatically. It should be stressed that
the requirements of the asperity smoothness, Eq.~(\ref{Smooth-2}),  and the
absence of the shadowing effect, Eq.~(\ref{Frenel}), are conventional in
solving problems of the wave diffraction at rough surfaces (see, e.g.,
Ref.~\cite{BassFuks}). The necessity of using them has not been overcome
till now.

It is instructive to note that in solving the diffraction problems the
condition of smallness of the so called Rayleigh parameter
$(k_z\sigma)^2$ is normally used. In the case of a single-channel strip,
Eq.~(\ref{1-chan}), the ratio $(\sigma/d)^2$ plays the role of this
parameter. The results presented herein are free of the above restriction.
Indeed, the ratio $(\sigma/d)^2$ was not thought to be small at any
step of handling the problem. Note that just the statistical identity and
complete correlation of the strip edges, Eq.~(\ref{xi-corr}), made it
feasible to bypass this restriction.

The main result of our work is revealing the remarkable sensitivity of the
interference effects in a single-mode waveguide to the intercorrelation
properties of the inhomogeneities of the opposite boundaries. To be
certain, it is sufficient to compare the localization length $L_0$,
obtained in Ref.~\cite{MakYur} for the conducting strip with only one
boundary rough, with the length $L_b$ from Eq.~(\ref{L_b}) of our paper.
In the former case $L_0\propto\sigma^{-2}$, whereas in ours
$L_b\propto\sigma^{-4}$. At first glance it would imply the CCB strips to
be more transparent for the electrons as against the junctions with
arbitrary asperities of the sides. However, it is not the case as a rule.
To illustrate this statement, assume the correlation
function ${\cal W}(x)$ of the asperities $\xi(x)$ as gaussian,
${\cal W}(x)=\exp\left(-x^2/2R_c^2\right)$. Then one can find the lengths
$L_0$ and $L_b$ related to each other as follows,

\begin{eqnarray}
L_0 / L_b \sim \left( \sigma/R_c\right)^2
\left( d/R_c \right)^2    \qquad
&\mbox{ if }& \qquad R_c/d \ll 1 \ (k_1 R_c\ll 1),
\nonumber \\
&& \label{L_0-L_b}  \\
L_0 / L_b \sim \left( \sigma / d \right)^2
\exp\left( k_1^2R_c^2  \right) \qquad
&\mbox{ if }& \qquad R_c/d \gg 1 \ (k_1 R_c\gg 1).
\nonumber
\end{eqnarray}
Note that in Eq.~(\ref{L_0-L_b}) the parameter $(\sigma/d)^2$  should be
thought small because the length $L_0$ was obtained in Ref.~\cite{MakYur}
under this assumption. It is evident from Eq.~(\ref{L_0-L_b}) that
the ratio $L_0 / L_b$ in both limiting cases is the product of a small
parameter by a large one. The parameters are such that the situation with
$L_0 \gg L_b$ is mostly realizable. Indeed, for the small-scale
asperities, when $k_1R_c\ll 1$ ($R_c/d\ll 1$), this is satisfied if the
slope $(\sigma/R_c)^2$ exceeds the small parameter $(R_c/d)^2$. In
the case of the large-scale asperities, i.e. $k_1 R_c\gg 1$
($R_c/d\gg 1$), the large exponent $(k_1R_c)^2$ must merely prevail the
logarithm $2\ln(d/\sigma)$.

The fact that localization lengths in single-mode strips with different
interboundary statistics of the inhomogeneities could deviate
significantly from one another can be explained, in our opinion, in a
following way. The localization length $L_0$ from Ref.~\cite{MakYur}
corresponds to the electron scattering by the effective potential

\begin{equation}
U_1 = {(\pi\hbar /d)^2 \over m}{\xi(x) \over d} \ ,
\label{U_1}
\end{equation}
which depends on just the asperity height $\xi(x)$ ($m$ is the electron
mass). In the CCB case, all the scattering potentials contain the gradient
$\xi'(x)$ instead of the function $\xi(x)$. Scattering by the potential
(\ref{U_1}) can be regarded as scattering by the asperity heights (or,
what is more precisely, by the waveguide width fluctuations). At the same
time, scattering by the potentials from Eq.~(\ref{G11-eq}) can be
interpreted as caused by the asperity slope fluctuations (or by the
waveguide bends). The strength of the by-height and by-slope scattering
depends on different parameters. Whereas the scattering from the potential
(\ref{U_1}) is governed by the Rayleigh parameter $(\sigma/d)^2$, the
by-slope scattering depends on the slope parameter $(\sigma/R_c)^2$.
Besides, not the least of the factors is the functional dependence of the
potentials on the random function $\xi(x)$. Indeed, the potential
(\ref{U_1}) is linear in $\xi(x)$ whereas the potentials from
Eq.~(\ref{G11-eq}) are quadratic in $\xi'(x)$. Thus, the distinction
between the scattering mechanisms in the waveguide with one boundary rough
and in the CCB strip brings the difference of the corresponding relaxation
lengths $L_0$ and $L_b$.

Another peculiarity of the electron scattering by the strongly correlated
identical rough edges is the necessity of taking into account the
`evanescent' waveguide modes, i.e. the non-propagating modes. These modes
are present in the last, i.e. the third, term in l.h.s. of the equation
(\ref{G11-eq}). As it is evident from the structure of the kernel
(\ref{K-ker}), this term governs intrachannel scattering of the
propagating mode with $n=1$ through interchannel transitions via the
virtual evanescent modes with $n\geq 2$. Those transitions contribute to the
expressions (\ref{L_f}), (\ref{L_b}) for the scattering lengths as much,
in order of magnitude, as the direct intramode scattering governed by the
potential $\hat{\cal V}(x)$ in Eq.~(\ref{G11-eq}). The conclusion
immediately follows that neglect of the evanescent modes in solving the
problems of waves and particles propagation in waveguides is not quite
correct in general. The present results demonstrate that this question
needs the special analysis every time it arises.

\appendix

% ------------------------------------------------------------------------
\section{Deriving the equation \protect\\
for the single-mode Green function}
\label{G11-K}
% ------------------------------------------------------------------------

In the case of the CCB waveguide, when Eqs.~(\ref{xi-corr}),
(\ref{corr=0}) hold,  the equation (\ref{Gnn'1}) for the mode Green
function $G_{nn'}(x,x')$ is represented as

\begin{equation}
\left[{\partial^2 \over \partial x^2} + k_n^2 + i0 -
\left({\pi n\over d}\right)^2 \hat{\cal V}(x) \right]G_{nn'}(x,x')
- {4\over d} \sum\limits_{m=1}^\infty A_{nm} \hat{\cal U}(x)G_{mn'}(x,x')
= \delta_{nn'}\delta(x-x') \, .
\label{Gnn-diff}
\end{equation}
This equation  with radiative boundary conditions at the strip ends
$x=\pm L/2$ is obviously equivalent to the Dyson-type integral equation,

\begin{eqnarray}
G_{nn^\prime}(x,x^\prime) &=& G_{n}^{(0)}(|x-x^\prime|) \delta_{nn^\prime}
+ \left(\frac{\pi n}{d}\right)^2 \int_{-L/2}^{L/2} dx_1
G_{n}^{(0)}(|x-x_1|) \hat {\cal V}(x_1) G_{nn^\prime}(x_1,x^\prime) +
\nonumber \\ \cr
&+& \frac{4}{d}\sum_{m=1}^{\infty}\int_{-L/2}^{L/2} dx_1
G_{n}^{(0)}(|x-x_1|) A_{nm}\,
\hat{\cal U}(x_1) G_{mn^\prime}(x_1,x^\prime).
\label{Dyson-n}
\end{eqnarray}
Here $G_{n}^{(0)}(|x-x^\prime|)$ is the unperturbed Green function
being the solution of Eq.~(\ref{Gnn-diff}) at
$\hat {\cal V}(x)\equiv \hat {\cal U}(x)\equiv 0$.

As $A_{nn}=0$, the equations (\ref{Gnn-diff}), (\ref{Dyson-n}) do not
contain the terms with $\hat {\cal U}(x)$ acting on
$G_{nn^\prime}(x,x^\prime)$. To account for this action we have to
substitute $G_{mn^\prime}(x,x^\prime)$ in the form (\ref{Dyson-n}) into
the last term in l.h.s. of Eq.~(\ref{Gnn-diff}). In doing so we obtain the
perturbative terms proportional to operators $\hat {\cal V}$,
$\ \hat {\cal U}\hat {\cal U}$, and $\hat {\cal U}\hat {\cal V}$.
Restricting ourselves, in view of the mildly sloping asperities
(\ref{smooth}), by only the perturbations quadratic in $\xi'(x)$ we
neglect the terms containing the product $\hat {\cal U}\hat {\cal V}$.
Then we get

\begin{eqnarray}
&& \left( \frac{\partial^2}{\partial x^2} +
k_n^2 + i0  \right) G_{nn^\prime}(x,x^\prime) -
\left(\frac{\pi n}{d}\right)^2 \hat {\cal V}(x)
G_{nn^\prime}(x,x^\prime) -
\nonumber \\ \cr
&& - \left(\frac{4}{d}\right)^2\sum_{m,m^\prime=1}^{\infty}
A_{nm}\,\hat{\cal U}(x) \int_{-L/2}^{L/2} dx_1 G_{m}^{(0)}(|x-x_1|)
A_{mm^\prime}\,\hat{\cal U}(x_1) G_{m^\prime n^\prime}(x_1,x^\prime) =
\nonumber \\ \cr
&& = \delta_{nn^\prime}\delta(x-x^\prime) +
\frac{4}{d}A_{nn^\prime}\,\hat{\cal U}(x)
G_{n^\prime}^{(0)}(|x-x^\prime|).
\label{GF-new}
\end{eqnarray}
It immediately follows from Eqs.~(\ref{GF-new}), (\ref{k_n}) that all
the off-diagonal Green functions $G_{nn^\prime}(x,x^\prime)$ with
$n\ne n'$ are small compared to the diagonal ones due to the second term in
r.h.s. of Eq.~(\ref{GF-new}).

Let us rewrite the equation (\ref{GF-new}) for the single-mode Green
function $G_{11}(x,x^\prime)$

\begin{eqnarray}
&& \left( \frac{\partial^2}{\partial x^2} +
k_1^2 + i0 \right) G_{11}(x,x^\prime) -
\left(\frac{\pi}{d}\right)^2 \hat {\cal V}(x) G_{11}(x,x^\prime) -
\nonumber \\ \cr
&& - \left(\frac{4}{d}\right)^2\sum_{m=2}^{\infty}
A_{1m}\,\hat{\cal U}(x) \int_{-L/2}^{L/2} dx_1 G_{m}^{(0)}(|x-x_1|)
A_{m1}\,\hat{\cal U}(x_1) G_{11}(x_1,x^\prime) -
\nonumber \\ \cr
&& - \left(\frac{4}{d}\right)^2\sum_{m,m^\prime=2}^{\infty}
A_{1m}\,\hat{\cal U}(x) \int_{-L/2}^{L/2} dx_1 G_{m}^{(0)}(|x-x_1|)
A_{mm^\prime}\,\hat{\cal U}(x_1) G_{m^\prime 1}(x_1,x^\prime)
= \delta(x-x^\prime).
\label{GF11-1}
\end{eqnarray}
The last term in l.h.s. of this equation has only the off-diagonal Green
functions with $m'\ge 2$ and can be consequently omitted. Thus we get from
Eq.~(\ref{GF11-1}) the asymptotically justified closed equation for
$G_{11}(x,x^\prime)$.

The mean value of the perturbative operator quadratic in $\hat {\cal U}$
in Eq.~(\ref{GF11-1}) differs from zero. The zero-mean-valued
operator necessary for the subsequent averaging over the random fields
can be obtained by merely subtracting the mean value of the original
operator from itself. In doing so we arrive at the equation,

\begin{eqnarray}
&& \left( \frac{\partial^2}{\partial x^2} +
k_1^2 + i0 \right) G_{11}(x,x^\prime) -
\left(\frac{\pi}{d}\right)^2 \hat {\cal V}(x) G_{11}(x,x^\prime) -
\left(\frac{4}{d}\right)^2 \int_{-L/2}^{L/2} dx_1 \hat K(x,x_1)
G_{11}(x_1,x^\prime) +
\nonumber \\ \cr
&& + \left(\frac{4}{d}\right)^2 \sum_{m=2}^{\infty} A_{1m}^2
\int_{-L/2}^{L/2} dx_1 \langle \hat{\cal U}(x) G_{m}^{(0)}(|x-x_1|) \,
\hat{\cal U}(x_1)\rangle G_{11}(x_1,x^\prime) = \delta(x-x^\prime).
\label{GF11}
\end{eqnarray}
Here the novel perturbation operator has occurred specified by the kernel
$\hat K(x,x^\prime)$, Eq.~(\ref{K-ker}). Besides, the additional, i.e. the
last, term has appeared in l.h.s. of Eq.~(\ref{GF11}). The detailed
analysis shows that this term gives rise to the small real renormalization
of the wavenumber $k_1$ and takes no effect on the relaxation processes.
This permits us to drop it from Eq.~(\ref{GF11}) and come directly to the
equation  (\ref{G11-eq}).

% ------------------------------------------------------------------------
\section{Formulation of the correlation relations \protect\\
for the space-averaged random fields}
\label{Fields}
% ------------------------------------------------------------------------

In Sec.~\ref{Model} we performed the averaging over the rapid phases and
arrived at the equations (\ref{PiGam}) in which the functions $\eta(x)$
and $\zeta_{\pm}(x)$ could be written as the sums

\begin{equation}
\eta(x) = S_V^+(x) + S_U^+(x) \, , \qquad
\zeta_-(x) = S_V^-(x) + S_U^-(x) \, , \qquad
\zeta_+(x) = \zeta_-^*(x) \, .
\label{etazeta-VU}
\end{equation}
The random fields $S_V^\pm(x)$ and $S_U^\pm(x)$ are associated
with the potentials $\hat{\cal V}(x)$ and ${\hat K}(x,x_1)$,

\begin{equation}
S_V^\pm(x) = \frac{1}{2k_1}\left(\frac{\pi}{d}\right)^2
\int_{x-l}^{x+l}\frac{dx^\prime}{2l}{\rm e}^{-ik_1x^\prime}
\hat {\cal V}(x^\prime)
{\rm e}^{\pm ik_1x^\prime} \, ,
\label{S_V-def}
\end{equation}

\begin{equation}
S_U^\pm(x) = \frac{1}{2k_1}\left(\frac{4}{d}\right)^2
\int_{x-l}^{x+l}\frac{dx^\prime}{2l}\int_{-L/2}^{L/2} dx_1
{\rm e}^{-ik_1x^\prime}\hat K(x^\prime,x_1){\rm e}^{\pm ik_1x_1} \, .
\label{S_U-def}
\end{equation}
Here the length $l$ is chosen arbitrary within the interval

\begin{equation}
\max\{k_1^{-1},R_c\} \ll l \ll \min\{L_1,L\} \, .
\label{l-def}
\end{equation}

In this Appendix we describe a way to obtain the correlation relations
(\ref{etazeta-corr}). We will demonstrate this with a simple example of
correlators of the fields $S_V^\pm(x)$ only. By substituting
$\hat{\cal V}(x)$ in the form (\ref{VUdef}) into Eq.~(\ref{S_V-def}) and
expressing $\xi(x)$ as the Fourier integral, we get

\begin{eqnarray}
S_V^\pm(x) &=& - \frac{1}{2k_1}\left(\frac{\pi}{d}\right)^2
\int_{-\infty}^{\infty}\frac{dq_x}{2\pi}(q_x \mp k_1)
\int_{-\infty}^{\infty}\frac{dq_x^\prime}{2\pi}(q_x^\prime-q_x)
\exp[i(q_x^\prime-k_1)x]\times
\nonumber \\ \cr
&\times&\frac{\sin[(q_x^\prime-k_1)l]}{(q_x^\prime-k_1)l}
\left[ \tilde\xi(q_x^\prime-q_x)\tilde\xi(q_x \mp k_1) -
\langle \tilde\xi(q_x^\prime-q_x)\tilde\xi(q_x \mp k_1)\rangle \right] \,,
\label{S-V}
\end{eqnarray}
with $\tilde\xi(q_x)$ being the Fourier transform of $\xi(x)$,

\begin{equation}
\tilde\xi(q_x) =
\int_{-L/2}^{L/2} dx \,
\xi(x) \exp(-iq_x x) \, .
\label{xi-qx}
\end{equation}
Assuming $\xi(x)$ to be the Gaussian random process we have the
correlation equalities for $\tilde\xi(q_x)$ resulting immediately
from Eq.~(\ref{xi-corr}),

\begin{equation}
\langle\tilde\xi(q_x)\rangle  =  0, \qquad
\langle\tilde\xi(q_x)\tilde\xi(q_x^\prime)\rangle=
\sigma^2 W(q_x)\Delta(q_x+q_x^\prime) \, .
\label{xi-qx-corr}
\end{equation}
Here $\Delta(q_x)$ indicates the `underlimiting' $\delta$-function,

\begin{equation}
\Delta(q_x) = \int_{-L/2}^{L/2} dx \exp(\pm iq_xx) =
\frac{\sin(q_xL/2)}{q_x/2}
\rightarrow 2\pi\delta(q_x) \, .
\label{Df}
\end{equation}

From Eqs.~(\ref{S-V}) and (\ref{xi-qx-corr}) we deduce the following
integral expression for the binary correlation function

\begin{eqnarray}
&& \langle S_V^\pm(x) S_V^\pm(x^\prime)\rangle =
\left(\frac{1}{2k_1}\right)^2 \left(\frac{\pi\sigma}{d}\right)^4
\int_{-\infty}^{\infty}\frac{dq_xdq_x^\prime
dq_x^{\prime\prime} dq_x^{\prime\prime\prime}}{(2\pi)^4}
(q_x \mp k_1) (q_x^\prime-q_x)
(q_x^{\prime\prime} \mp k_1)
(q_x^{\prime\prime\prime}-q_x^{\prime\prime})\times
\nonumber \\ \cr
&&\times
W(q_x \mp k_1)W(q_x^\prime - q_x)
\exp[i(q_x^\prime-k_1)x +i(q_x^{\prime\prime\prime}-k_1)x^\prime]
\frac{\sin[(q_x^\prime-k_1)l]}{(q_x^\prime-k_1)l}
\frac{\sin[(q_x^{\prime\prime\prime}-k_1)l]}
{(q_x^{\prime\prime\prime}-k_1)l}
\times \nonumber \\ \cr
&&\times
\Delta(q_x^{\prime\prime\prime} + q_x^\prime \mp 2k_1)
\left[ \Delta(q_x^{\prime\prime} + q_x \mp 2k_1) +
\Delta(q_x^{\prime\prime} + q_x^\prime - q_x \mp k_1)\right].
\label{SS2}
\end{eqnarray}
The integrand of Eq.~(\ref{SS2}) contains three types of sharp
functions. The first is $\Delta(q_x)$ with variation scale
$q_x \sim L^{-1}$, the second, $W(q_x)$, varies at $q_x \sim R_c^{-1}$,
and the third-type functions are those of the form $\sin(q_x l)/q_x l$.
Owing to Eq.~(\ref{l-def}), the function $\Delta(q_x)$ is the sharpest in
the integrand. With its aid we take the integrals over $q''_x$ and
$q'''_x$. The $q'_x$-integral is evaluated through the third-type sharp
functions. In such a way we obtain the formula

\begin{eqnarray}
\langle S_V^\pm(x)S_V^\pm(x^\prime)\rangle &=&
\frac{3\mp 1}{8k_1^2}\left(\frac{\pi\sigma}{d}\right)^4
\int_{-\infty}^{\infty}\frac{dq_x}{2\pi}
(q_x \mp k_1)^2 W(q_x \mp k_1)\times
\nonumber \\ \cr
&\times&\left[(q_x-k_1)^2 W(q_x-k_1) +
(q_x+k_1\mp 2k_1)^2 W(q_x+k_1\mp 2k_1)\right] \times
\nonumber \\ \cr
&\times&\exp[-i(k_1\mp k_1)(x+x^\prime)]
F_l^\pm(x-x^\prime) \, .
\label{SS4}
\end{eqnarray}

The functions $F_l^\pm(x)$ in Eq.~(\ref{SS4}) are

\begin{equation}
F_l^+(x) =
\frac{1-|x|/2l}{2l}\Theta(2l-|x|) \, ,
\qquad
F_l^-(x) =
\frac{\sin[4(1-|x|/2l)k_1l]}{8k_1l^2}\Theta(2l-|x|) \, .
\label{F_l-def}
\end{equation}
The function $F_l^+(x)$ is sharp within the scales $L_1$ and $L$ with mean
value equal to unity,

\begin{equation}
\int_{-\infty}^{\infty}dx\,F_l^+(x) = 1.
\label{F_l-int}
\end{equation}
Thus this function can be replaced by the $\delta$-function in the
correlator $\langle S_V^+(x)S_V^+(x^\prime)\rangle$. At the same time,
the function $F_l^-(x)$ is integrally small in the parameter
$(k_1l)^{-2}\ll 1$ and, consequently, is allowed to be put zero.
Taking this into account we get the final expressions for the correlators
(\ref{SS2}), with the accuracy prescribed by the conditions (\ref{l-def}),

\begin{eqnarray}
\langle S_V^+(x)S_V^+(x^\prime)\rangle &=&
L_f^{-1}\{VV\} \delta(x-x^\prime) \, ;
\nonumber \\[-6pt]
&& \label{SSV} \\[-6pt]
\langle S_V^-(x)S_V^-(x^\prime)\rangle &=& 0 \, .
\nonumber
\end{eqnarray}
Here the notation $L_f\{VV\}$ stands for the electron relaxation
length conditioned by the potential $\hat{\cal V}(x)$ and corresponds to
the forward electron scattering. From Eq.~(\ref{SS4}) it follows that

\begin{equation}
L_f^{-1}\{VV\} = \frac{1}{2k_1^2}\left(\frac{\pi\sigma}{d}\right)^4
\int_{-\infty}^{\infty}\frac{dq_x}{2\pi}q_x^4W^2(q_x) \, .
\label{L_f-VV}
\end{equation}

Performing the analogous calculations for the correlators
$\langle S_V^\pm(x) S_V^{\pm*}(x^\prime)\rangle$ we find, with the same
accuracy,

\begin{eqnarray}
\langle S_V^+(x)S_V^{+*}(x^\prime)\rangle &=&
L_f^{-1}\{VV\} \delta(x-x^\prime) \, ;
\nonumber \\[-6pt]
&& \label{SSV*} \\[-6pt]
\langle S_V^-(x)S_V^{-*}(x^\prime)\rangle &=&
L_b^{-1}\{VV\} \delta(x-x^\prime) \, .
\nonumber
\end{eqnarray}
Here $L_b^{-1}\{VV\}$ is the backward-scattering relaxation length
specified by the expression

\begin{equation}
L_b^{-1}\{VV\} = \frac{1}{2k_1^2}\left(\frac{\pi\sigma}{d}\right)^4
\int_{-\infty}^{\infty}\frac{dq_x}{2\pi}(k_1^2-q_x^2)^2
W(k_1-q_x)W(k_1+q_x) \, .
\label{L_b-VV}
\end{equation}

Calculation of all the remaining correlators of the functions
(\ref{S_V-def}), (\ref{S_U-def}), necessary for obtaining the correlation
relations for the fields $\eta(x)$ and $\zeta_{\pm}(x)$, can be done in a
perfectly similar way. Minor additional complications are connected with
the unwieldy structure of the kernel $\hat K(x^\prime,x_1)$ only,
Eq.~(\ref{K-ker}). They can be easily overcome having in mind the
weak-scattering conditions (\ref{two-scale}). The result is given by
Eqs.~(\ref{etazeta-corr}) -- (\ref{L_b}).

% ------------------------------------------------------------------------


\begin{references}

\bibitem{TrAsh}
Trivedi N and Ashcroft N W 1988
{\it Phys. Rev.} B {\bf 38} 12298

\bibitem{MakYur}
Makarov N M and Yurkevich I V 1989
{\it Zh. Eksp. Teor. Fiz} {\bf 96} 1106
(Engl. Transl. {\it Sov. Phys.--JETP} {\bf 69} 628)

\bibitem{KL}
Kierul J and Ledzion J 1991
{\it Phys. Stat. Sol.} (A) {\bf 128} 117

\bibitem{TakFer}
Takagaki Y and Ferry D K 1992
{\it J. Phys.: Condens. Matter} {\bf 4} 10421

\bibitem{Kun}
Kunze Chr 1993
{\it Solid State Commun.} {\bf 87} 359

\bibitem{KoKr}
Kozub V I and Krokhin A A 1993
{\it J. Phys.: Condens. Matter} {\bf 5} 9135

\bibitem{MeyStep}
Meyerovich A E and Stepaniants S 1995
{\it Phys. Rev.} B {\bf 51} 17116

\bibitem{MakMor}
Makarov N M, Moroz A V and Yampol'skii V A 1995
{\it Phys. Rev.} B {\bf 52} 6087

\bibitem{LunKr}
Luna-Acosta G A, Krokhin A A, Rodriguez M A and
Hernandez-Tejeda P H 1996
{\it Phys. Rev.} B {\bf 54} 11410

\bibitem{Kubo}
Kubo R 1957
{\it J. Phys. Soc. Japan} {\bf 12} 570

\bibitem{FishLee}
Fisher D S and Lee P A 1981
{\it Phys. Rev.} B {\bf 23} 6851

\bibitem{Edwards}
Edwards S F 1958
{\it Phil. Mag.} {\bf 3} 1020

\bibitem{Abrik}
Abrikosov A A and Ryzhkin I A 1978
{\it Adv. Phys.} {\bf 27} 147

\bibitem{Efetov}
Efetov K B 1983
{\it Adv. Phys.} {\bf 32} 53

\bibitem{Bogol}
Bogolyubov N N and Mitropolskii Y A 1974
{\it Asymptotic methods in a theory of non-linear oscillations},
4-th Ed.
(Moscow: Nauka, {\it in Russian})

\bibitem{Papanic}
Asch M, Kohler W, Papanicolaou G, Postel M and White B 1991
{\it SIAM Rev.} {\bf 33} 519

\bibitem{Mehler}
Mehler F G 1881
{\it Math. Ann.} {\bf 18} 161

\bibitem{Fock}
Fock V A 1943
{\it Doklady AN SSSR} {\bf 39} 253

\bibitem{LifGredPas}
Lifshits I M, Gredeskul S A and Pastur L A 1988
{\it Introduction to the Theory of Disordered Systems}
(New York: Wiley)

\bibitem{BassFuks}
Bass F G and Fuks I M 1979
{\it Wave Scattering from Statistically Rough Surfaces}
(New York: Pergamon)

\end{references}
\end{document}